# Assessing Strong String Stability of Constant Spacing Policy under Speed Limit Fluctuations

Sina Arefizadeh[1], Aria Hasanzadezonuzy[2], Alireza Talebpour[3]*, Srinivas Shakkottai[4]

*Abstract*— The speed limit changes frequently throughout the transportation network, due to either safety (e.g., change in geometry) or congestion management (e.g., speed harmonization systems). Any abrupt reduction in the speed limit can create a shockwave that propagates upstream in traffic. Dealing with such an abrupt reduction in speed limit is particularly important while designing control laws for a platoon of automated vehicles from both stability and efficiency perspectives. This paper focuses on Adaptive Cruise Control (ACC) based platooning under a constant spacing policy, and investigates the possibility of designing a controller that ensures stability, while tracking a given target velocity profile that changes as a function of location. An ideal controller should maintain a constant spacing between successive vehicles, while tracking the desired velocity profile. The analytical investigations of this paper suggest that such a controller does not exist.

## I. INTRODUCTION

Changing the speed limit is a common strategy to deal with safety concerns (e.g., change in the roadway geometry, school zones, adverse weather condition, etc.) and to mitigate congestion (e.g., speed harmonization systems [1]). A sudden change in speed limit can result in shockwave formation and propagation, which can potentially change the traffic flow regime and cause flow breakdown. While the current practice in transportation systems adjusts the speed limit at fixed locations throughout the roadway network, the introduction of connected vehicles technology provides the opportunity to enhance safety and reduce congestion by providing speed limit information at any arbitrary location along the roadway network. This technology also provides the necessary information to identify shockwaves at the onset of formation.

However, the introduction of connected vehicle technology, coupled with potential speed changes in arbitrary locations introduces new challenges for operating platoons of automated vehicles. In the past two decades, many applications of automated driving, including Adaptive Cruise Control (ACC) and Cooperative Adaptive Cruise Control (CACC), have been developed ([2],[3],[4]). These platooning applications are designed to take advantage of low reaction times in automated vehicles to maximize roadway capacity and mitigate congestion. Two main approaches have been adopted in previous studies, namely, (i) constant spacing policy, and (ii) constant headway policy. The constant spacing policy is focused on keeping a constant space between the vehicles, while the time headway policy is focused on keeping a fixed time-headway between vehicles[5, 6]. Platoon stability is the key factor in the success of these platooning applications. In fact, the individual vehicle control logic should ensure that the entire platoon stays stable while dealing with a sudden speed drop.

[1] Dept. of Civil Engineering, Texas A&M University, Email: sinaarefizadeh@tamu.edu.
[2] Dept. of Electrical and Computer Engineering, Texas A&M University, Email: azonuzy@tamu.edu.
[3] Dept. of Civil Engineering, Texas A&M University, (Corresponding Author) Email: atalebpour@tamu.edu, Phone: 979.845.0875.
[4] Dept. of Electrical and Computer Engineering, Texas A&M University, Email: sshakkot@tamu.edu.

To analyze the stability of a platoon facing this situation, this study introduces a new definition, "strong string stability," or in short, "strong stability," (a generalization of the so-called "exponential string stability"[7] based on two notions of error associated with the platoon. As Fig. 1 illustrates, the actual distance between vehicle $i+1$ and vehicle $i$, as compared to the target following distance $r$, defines localized error between the two, referred to as $e_i$. Since each pair of vehicles experiences an error, the location of vehicle $i+1$ with respect to an imaginary lead vehicle traveling at constant speed has an error from its target location by the sum of all the pairwise errors of vehicles ahead of it. Such a sum over errors can be thought of as the platoon-level error as measured at vehicle $i+1$.

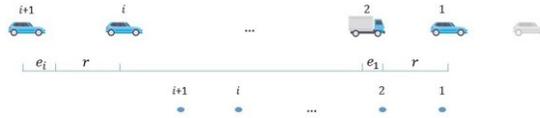

Fig 1. Local and global reference locations. $e_i$ is error term with respect to local target location. Blue dots represent global target locations with respect to the leading vehicle. the shadowed vehicle in gray is the imaginary vehicle.

A definition of a strong stable platoon that accounts for both the local and platoon-level errors should have the following characteristics: (i) all vehicles should be at their desired reference point with respect to vehicle ahead asymptotically in time, and (ii) all vehicles should be within a bounded error from their desired target location (with respect to the imaginary lead vehicle) asymptotically in time. The first characteristic is a necessary condition for a system to follow a constant spacing policy, which is our key requirement to ensuring platoon safety. The second characteristic guarantees a finite delay in vehicles travel time in the platoon. Thus, a strong stable system guarantees the asymptotic performance of the string of vehicles, since all vehicles would be at their reference point. Such a condition enables one to accurately predict traffic flow dynamics. As a result, specific densities and flows could be guaranteed at the steady-state condition by choosing the spatial velocity profile appropriately.

The main objective of this study is to investigate the possibility of ensuring strong stability under the constant spacing policy (where each autonomous vehicle tries to keep a pre-determined distance from its leader) in the presence of a sudden speed drop. The main contribution of this paper is to show that regardless of the individual vehicles' control law, a constant spacing policy cannot be maintained in a strong stable manner in the presence of a sudden speed drop. As a result, the platooning based on constant spacing policy cannot result in a stable traffic flow since an exponentially string stable (i.e. strong stable) control scheme does not exist based on constant spacing in the presence of a speed drop. Note that ensuring traffic flow stability is critical for the success of any platooning strategy. It is noteworthy that few studies investigated the possibility of ensuring string stability in a platoon based on the constant spacing policy (see [8-11] for some examples). Ioannou and his colleagues studied string stability for particular type of controllers such as PID and adoptive control based on the constant spacing and other platooning strategies [8, 9]. Ioannou and Chien [10] showed that V2V communications is the necessary condition to ensure string stability in a platoon under the constant spacing policy. Later, Jovanovic and Bamieh [11] presented several issues associated with the common approaches in vehicular platooning. Moreover, they presented a formulation of optimal control for platooning problem that we will discuss it later in the paper. However, the main contribution of current study is to show that platooning based on the constant spacing policy cannot result in a stable traffic flow, since a strong stable control scheme does not exist based on the constant spacing in the presence of a speed drop.

The remainder of this paper is organized as follows: the next section presents the formal problem statement and key definitions. Then, the control model is introduced along with the proof that if there exists a control system guaranteeing a constant relative distance of successive vehicles, it is not able to track a speed drop. This section is followed by a discussion on the findings of this study. Finally, significance and contributions and concluding remarks and future research needs close the paper.

## II. Significance and contributions

As it is mentioned earlier in the paper, previous studies [10, 11] presented some result in the similar cases. In these papers considering a spatially invariant problem and a second ordered model for vehicles in the platoon, authors illustrated that exponential string stability is impossible to achieve when only information on relative position errors are available. In detail, at part 2), subsection B of the section III of [10], where reaching the cruise speed is the goal, authors mentioned " The lack of information about the actions of the lead vehicle causes a disturbance amplification in the value of deviation constant spacing, velocities, and accelerations of the following vehicles". Additionally, at subsection C of section II of [11], defining the optimal control formulation and introducing the quadratic cost function, for the case of approaching the cruise speed $v_d$, authors proved that considering relative position errors of successive vehicles will not result in exponentially stable string as the size of the platoon goes to infinity. As a result, the formulation is ill-posed. However, later in this paper at section IV and V, authors modified the formulation by introducing the absolute error terms to the formulation and proved that in this case problem become well-posed and achieving an exponentially stable string is possible, which is equivalent to the case of providing information from the leading vehicle.

In the current research, after justifying why spatial variation problem is important, we showed that when the velocity profile is not constant, achieving exponentially string stable platoon using constant spacing policy in the presence or absence of V2I is impossible.

The analysis in the paper is independent of the level of information provided and it is only based on the system behavior around the potential equilibrium of the system. In other words, having information about relative position errors and velocity errors with respect to the velocity profile is sufficient to induce impossibility of the exponential string stability and extra information regarding absolute position errors has not been used throughout the analytical argument of the current paper. However, tracking velocity profile beside relative position errors may imply such information indirectly.

In sum, the current paper can be considered as an extension of the result of the constant spacing case studied in the [11], where the problem is spatially variant.

## III. Problem Statement

Consider the desired velocity profile of Fig. 2 that shows a sudden drop in the desired velocity. Assume that vehicle $i$ reaches the speed drop location at time $i.T$, where $T$ is the time difference between the arrival of two consecutive vehicles at the speed drop location. All vehicles are supposed to follow this velocity profile, where $v_0$ is the velocity of the string before the drop location, and $(1-p).v_0$ is the velocity of the string after the drop location. $p.v_0$ is the amount of drop in velocity at the speed drop location, where $0 \leq p \leq 1$. The objective is to ensure that all vehicles follow the profile presented in Fig. 2 and maintain a fixed spacing with respect to the next vehicle in the string.

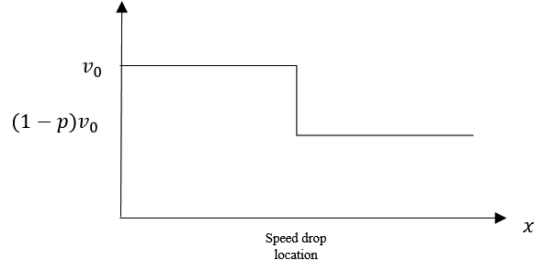

Fig 2. Desired Speed Profile.

As discussed in the Introduction, a useful set up is to consider an imaginary vehicle with a fixed velocity $v_0$, which provides a reference location that other vehicles can track. Hence, the rest of the vehicles follow the trajectory of this imaginary vehicle with a delay due to speed drop within a time shift of T. Fig. 3 illustrates the trajectories of the imaginary and actual vehicles. We indicate the imaginary lead vehicle by index 0, and real vehicles following it starting with index 1.

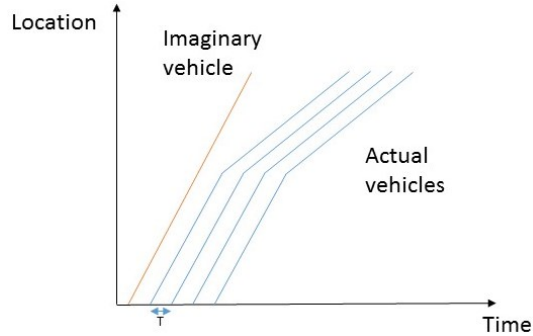

Fig 3. Imaginary vs. Actual Vehicle Trajectories.

String stability is a concept defined to ensure a bounded error in states of an interconnected system throughout the time and space, and preventing the errors from accumulating [12]. In other words, according to this definition the error for vehicle $i$, $e_i$, which is defined as the error term between reference point of vehicle $i$ (i.e., desired position) and its actual position with respect to vehicle ahead, should be greater than or equal to the error for vehicle $i + 1$, $e_{i+1}$, where vehicle $i + 1$ is the immediate follower. Therefore, even when a shockwave is produced by a disturbance (e.g., by a sudden brake), the system would still be called string stable if the wave propagates upstream with the same disturbance rate. In this extreme case, the produced error remains in the string, and vehicles deviate from their reference points. The collective movement of vehicles might become unpredictable.

Exponential string stability is another concept introduced to ensure string stability and zero error for all vehicles asymptotically. (A string of vehicles is called exponential string stable if (1) $\left\|\frac{E_i(jw)}{E_{i-1}(jw)}\right\| \leq 1$ for every $i \in N$, $w \in \mathbb{R}$

[13], and (2) $\lim_{s \to 0} s.E_i = 0$ for every $i \in N$ [7], where $E_i(s)$ denotes the error term for vehicle $i$ in Laplace domain.) It is not difficult to show that exponentially string stability, when error propagation rate is strictly decreasing, results in our definition of strong stability. In a Lemma introduced at the end of Section III, this result is shown for homogenous platoons of vehicles. We first introduce the key definitions, and then proceed to evaluate the objective. The following are the key definitions in this paper:

**Definition 1.** Local stability for vehicle $i$ is defined as follows:
$$\lim_{t \to \infty} e_i(t) = \lim_{s \to 0} s.E_i = 0,$$
where $e_i$ is the error term in location in time and $E_i$ denotes the Laplace of $e_i$.

**Definition 2.** A platoon is strong stable if local stability holds for each vehicle and:
$$\lim_{t \to \infty} \sum_{i=1}^{\infty} e_i(t) < M < \infty.$$

An immediate consequence of violating strong stability is that the vehicle deviates infinitely from its reference point associated with the leading vehicle. Therefore, if its reference point is assumed to follow a specific velocity profile, the actual vehicle will follow the velocity profile with an infinite delay, which is not desirable.

IV. CONTROL MODEL

Let us consider the control model illustrated in Fig. 4 to evaluate the vehicles' ability to follow the desired velocity profile. This is a generic control low and no specific controller is considered. Moreover, it addresses both the constant spacing as well as velocity-tracking requirement in our problem. We consider an imaginary leading vehicle traveling at a constant velocity, $v_0$. Vehicle $i$ is supposed to reduce its speed to $p.v_0$ ($0 \leq p \leq 1$) at time $i.T$, while maintaining a constant spacing $r$ between vehicles. The feedback loop including controller $k_p$ is designed to keep a constant distance between two successive vehicles. In order to address the tracking problem, let us consider another controller that regulates the velocity profile of each vehicle. A feedback loop, including controller $K_l$, tracks the desired velocity profile by means of tracking the imaginary leader vehicle's position with a predefined delay resulting in tracking the desired velocity profile. The predefined delay should be of the form of ramp to guarantee a certain step delay in velocity.

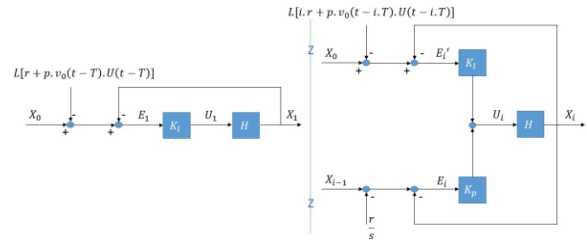

Fig4. Control Model.

For each vehicle, controlling two output using one input in the linear control framework is impossible (see Linear Control Theory, Lemma 13.6 [14]). Hence, we need a combination of two controllers which are $K_l, K_p$.

In the control model of Fig4:

r: constant spacing,

p: amount of the sudden drop,

$v_0$: initial speed before the speed drop location,

i.T: time when vehicle $i$ reaches the drop location,

U(.): unit step function,

$E_i$: error in Laplace domain between spatial positioning of vehicle $i$ and its actual location with respect to the next vehicle in the platoon,

$K_l, K_p$: leading and predecessor controller,

$X_0(s)$: Target location of imaginary reference vehicle in the Laplace-domain,

$X_1(s)$: actual leading vehicle position in the Laplace domain,

$E'_i$: difference between the actual location of vehicle $i$ and its imaginary reference point,

H: Dynamics of the vehicle,

L[.]: Laplace of the function inside the bracket.

Our goal is to identify the conditions for strong stability in a platoon of autonomous vehicles using the above control system under the speed drop scenario.

Proceeding in the manner of Shaw and Hedrick [13], we define the control laws as follows:

$$U_1 = E_1 . K_l, \qquad (1)$$

$$U_i = E'_i . K_l + E_i . K_p. \qquad (2)$$

The following equations can be derived based on the controller (see Fig. 4 for more details):

$$X_1 = E_1 . K_l . H, \qquad (3)$$

$$E_1 = \frac{X_0 - \frac{r}{s} - \frac{pv_0}{s^2} e^{sT}}{1 + K_l . H}. \qquad (4)$$

For all other vehicles ($i \geq 2$), we have:

$$X_i = K_p . E_i . H + K_l . H \left( X_0 - i.\frac{r}{s} - \frac{pv_0}{s^2} e^{s.i.T} - X_i \right),$$

$$X_i = \frac{K_p . H}{(1 + K_l . H)} . E_i + \frac{K_l H}{(1 + K_l . H)} . \left( X_0 - i.\frac{r}{s} - \frac{pv_0}{s^2} e^{s.i.T} \right). \qquad (5)$$

Consequently, the following equations can be derived based on the definition of $E_2$:

$$E_2 = X_1 - X_2 - \frac{r}{s},$$

$$E_2 = \frac{X_0 - \frac{r}{s} - \frac{pv_0}{s^2} e^{sT}}{1 + K_l . H} . K_l . H - \frac{K_p . H}{(1 + K_l . H)} . E_2 + \frac{K_l H}{(1 + K_l . H)} . \left( X_0 - 2.\frac{r}{s} - \frac{pv_0}{s^2} e^{2.s.T} \right) - \frac{r}{s},$$

$$E_2 = \frac{-K_l . H}{[1 + (K_l + K_p) H]} . \left( \frac{pv_0}{s^2} e^{s.T} - \frac{pv_0}{s^2} e^{2s.T} \right) - \frac{\frac{r}{s}}{[1 + (K_l + K_p) H]}. \qquad (6)$$

Consequently, the following recursive relation can be derived for $E_i$:

$$E_i = \frac{K_p.H}{[1+(K_l+K_p)H]} \cdot E_{i-1} + \frac{K_l.H}{[1+(K_l+K_p)H]} \cdot \left(\frac{pv_0}{s^2}e^{i.s.T} - \frac{pv_0}{s^2}e^{(i-1).s.T}\right) - \frac{\frac{r}{s}}{[1+(K_l+K_p)H]}. \tag{7}$$

**Theorem 1.** If there exists a linear controller guaranteeing: (i) constant relative distance of successive vehicles, (ii) tracking of a velocity profile as a function of location that takes the form of a step function, and (iii) strong stability, it should satisfy the following equation:

$$\lim_{x\to 0, n\to\infty}\left\{-\frac{K_p.H}{1+K_l.H}\cdot\frac{x.E_n}{n} - \frac{K_l.H}{1+K_l.H}\cdot\frac{pv_0}{n.x}(e^{x.T}-e^{n.x.T}) - \frac{n-1}{n}\cdot\frac{r}{1+K_l.H}\right\} = 0.$$

**Proof.** Let us define: $\tilde{E}_n = \sum_{i=2}^n E_i$, the following equation can be obtained from (7):

$$\tilde{E}_n - E_2 = \frac{K_p.H}{[1+(K_l+K_p)H]}\cdot\tilde{E}_n - \frac{K_p.H}{[1+(K_l+K_p)H]}\cdot E_n - \frac{K_l.H}{[1+(K_l+K_p)H]}\cdot\left(\frac{pv_0}{s^2}e^{2.s.T} - \frac{pv_0}{s^2}e^{n.s.T}\right) - \frac{\frac{(n-2)r}{s}}{[1+(K_l+K_p)H]}. \tag{8}$$

Simplifying (8), we have:

$$\frac{1+K_l.H}{[1+(K_l+K_p)H]}\cdot\tilde{E}_n = -\frac{K_p.H}{[1+(K_l+K_p)H]}\cdot E_n - \frac{K_l.H}{[1+(K_l+K_p)H]}\cdot\left(\frac{pv_0}{s^2}e^{s.T} - \frac{pv_0}{s^2}e^{n.s.T}\right) - \frac{\frac{(n-1)r}{s}}{[1+(K_l+K_p)H]}. \tag{9}$$

One can multiply (9) by $\frac{[1+(K_l+K_p)H]}{1+K_l.H}\cdot\frac{s}{n}$ (for any arbitrary $n>0$) to obtain

$$\frac{s.\tilde{E}_n}{n} = -\frac{(K_p.H)}{(1+K_l.H)}\cdot\frac{s}{n}\cdot E_n - \frac{(K_l.H)}{(1+K_l.H)}\cdot\frac{s}{n}\cdot\left(\frac{pv_0}{s^2}e^{s.T} - \frac{pv_0}{s^2}e^{n.s.T}\right) - \frac{(n-1)r}{n.(1+K_l.H)}. \tag{10}$$

Since (10) is valid for every arbitrary $n$, we have

$$\lim_{s\to 0, n\to\infty}\frac{s.\tilde{E}_n}{n} = \lim_{s\to 0, n\to\infty}\left\{-\frac{(K_p.H)}{(1+K_l.H)}\cdot\frac{s}{n}\cdot E_n - \frac{(K_l.H)}{(1+K_l.H)}\cdot\frac{s}{n}\cdot\left(\frac{pv_0}{s^2}e^{s.T} - \frac{pv_0}{s^2}e^{n.s.T}\right) - \frac{(n-1)r}{n.(1+K_l.H)}\right\}. \tag{11}$$

Now, (11) should be satisfied for every particular path of $s\to 0$ in the complex plane. Let us consider $s = x + y*j$ in the direction of $x + 0*j$, which passes through origin. Then we have

$$\lim_{x\to 0, n\to\infty}\frac{x.\tilde{E}_n}{n} = \lim_{x\to 0, n\to\infty}\left(-\frac{(K_p.H)}{(1+K_l.H)}\cdot\frac{x}{n}\cdot E_n - \frac{(K_l.H)}{(1+K_l.H)}\cdot\frac{1}{n}\cdot\left(\frac{pv_0}{x}e^{x.T} - \frac{pv_0}{x}e^{n.x.T}\right) - \frac{(n-1)}{n.(1+K_l.H)}r\right). \tag{12}$$

Assuming that the platoon is strong stable, we must have $\lim_{t\to\infty, n\to\infty}\tilde{e}_n(t) = M < \infty$. This implies that, according to the final value theorem, $\lim_{s\to 0, n\to\infty} s.\tilde{E}_n(s) = M$.

Therefore, the left hand side of (12) can be rewritten as follows:

$$\lim_{x\to 0, n\to\infty}\frac{x.\tilde{E}_n}{n} = \lim_{,n\to\infty}\frac{1}{n}\cdot\lim_{x\to 0, n\to\infty} x.\tilde{E}_n = 0.$$

Hence, a necessary condition for the stability is

$$\lim_{x\to 0, n\to\infty}\left\{-\frac{(K_p.H)}{(1+K_l.H)}\cdot\frac{x}{n}\cdot E_n - \frac{(K_l.H)}{(1+K_l.H)}\cdot\frac{1}{n}\cdot\left(\frac{pv_0}{x}e^{x.T} - \frac{pv_0}{x}e^{n.x.T}\right) - \frac{(n-1)r}{n.(1+K_l.H)}\right\} = 0. \tag{13}$$

Note that for all paths of $x \to 0, n \to \infty$, (13) should be satisfied. Thus, (13) provides a necessary condition for the stability of platoons of vehicles. ∎

The next question is whether we can construct a controller than can satisfy the necessary conditions that we have derived. We first define $f_0(s)$ as polynomial of $s$ such that $f_0(s)$ is non-singular ($s$ and $f_0(s)$ are coprime). Then we have the following corollary.

**Corollary.** There is no controller satisfying the conditions of Theorem 1

**Proof.** Substituting $f_0(s)$ into (13), there are some polynomials of $f_0^1(s), f_0^2(s), f_0^3(s)$ and $a, b, c \in Z$ such that:

$$\lim_{x \to 0, n \to \infty} \left\{ \frac{x^a}{n} \cdot \frac{f_0^2(x)}{f_0^1(x)} \cdot (x.E_n) + x^b \cdot \frac{f_0^3(x)}{f_0^1(x)} \cdot \frac{pv_0}{n} (e^{x.T} - e^{n.x.T}) + x^c \cdot \frac{n-1}{n} \cdot \frac{r}{f_0^1(x)} \right\} = 0. \tag{14}$$

If $\min\{a, b, c\}$ is a negative number, we can multiply (9) by $s^{-\min\{a,b,c\}}$ and rewrite (10) through (13) so that the new values of $a, b,$ and $c$ are non-negative. Hence, without any loss of generality, let us consider these parameters as non-negative numbers. Rearranging (14), we have:

$$\lim_{x \to 0, n \to \infty} \left\{ \frac{x^a}{n} \cdot \frac{f_0^2(x)}{f_0^1(x)} \cdot (x.E_n) + x^c \cdot \frac{n-1}{n} \cdot \frac{r}{f_0^1(x)} \right\} = - \lim_{x \to 0, n \to \infty} x^b \cdot \frac{f_0^3(x)}{f_0^1(x)} \cdot \frac{pv_0}{n} (e^{x.T} - e^{n.x.T}). \tag{15}$$

The left hand side of (15) exists and equals zero; therefore, the right hand side must exist and equal zero for every path in which $x \to 0, n \to \infty$. Let us consider the path of $x = \frac{1}{\sqrt{n}}$. Then

$$\lim_{n \to \infty} \frac{1}{n^{\frac{b}{2}+1}} \cdot e^{T.\sqrt{n}} = \lim_{n \to \infty} \frac{e^{T.\sqrt{n}}}{e^{(\frac{b}{2}+1)*\ln(n)}} = \infty. \tag{16}$$

Therefore, (16) yields a contradiction, since necessary condition of (13) does not hold. ∎

**Remark 1.** The problem of determining local stability (sometimes called plant stability) in control systems is typically analyzed by the Hurwitz criteria for continuous systems. Due to the fact that the term $e^{isT}$ appears in our analysis, addressing stability issues using such an approach is challenging. Therefore, this paper adopts a more basic analysis based on the final value theorem.

The combination of above theorem and corollary indicates that a controller that can maintain a fixed spacing among the vehicles in the platoon, while following a sudden speed drop cannot exist. Below is an illustration of a more specific scenario.

**Lemma.** For a homogenous string of vehicles, exponential string stability with strictly decreasing error propagation rate results in strong stability.

**Proof.** Since at exponential string stability local (plant) stability holds for every single vehicle in the string we have:

$$\lim_{s \to 0} s. E_n(s) = \lim_{t \to 0} e_n(t). \tag{17}$$

Since $\frac{\|E_{i+1}\|}{\|E_i\|} = P < 1$

$$\frac{1}{1-p}\|E_1\| = \sum_{i=1}^{\infty}\|E_i\|, \text{ and} \qquad (18)$$

$$\lim_{s\to 0}\|s\|\cdot\frac{1}{1-p}\|E_1\| = \lim_{s\to 0}\|s\|\cdot\sum_{i=1}^{\infty}\|E_i\| = \lim_{s\to 0}\sum_{i=1}^{\infty}\|s\cdot E_i\|. \qquad (19)$$

Considering local stability, (17) should be zero. Therefore, if each side of (19) equals zero, the other side should be zero as well. It should be noted that left hand side of (19) determines total error in the string with error propagation rate less than one. Thus, the right hand side of (19) shows total error in the string, which equals zero if and only if all vehicles asymptotically take place at their predefined reference point. Such a criterion implies strong stability. ∎

## V. DISCUSSION

In the linear control framework, equilibrium points play a key role. The asymptotic behavior of a stable system with zero steady-state error can be determined by studying the equilibrium. In order to have a generic control model, we assumed there is a feedback loop, including an arbitrary controller that guarantees constant spacing asymptotically, and another feedback loop, including an arbitrary controller, that addresses the speed tracking (following the step in the velocity profile). Since it is impossible to control two output (i.e., error in relative distance and error in velocity) using one input in linear control, the main controller should be a combination of these controllers. Therefore, we proposed the generic control model in Fig. 4. For a constant velocity profile, it results in leader-predecessor following control strategy that its stability for a constant velocity profile was addressed [15].

In this research, we claimed that it is impossible to simultaneously achieve all the following for the constant spacing policy, while tracking a velocity profile that has a sudden drop:

1. Strong stability,
2. Local stability,
3. A linear controller.

Violating either one of strong stability, and local stability has some undesirable consequences. First, since summation of errors in relative distance of all vehicles downstream appears in the error term for tracking velocity profile, violating the strong stability causes a vehicle to track velocity profile with an infinite delay. Such an infinite delay will cause infinite travel time. Second, violating local stability means constant relative distance is impossible. Considering all the above findings and discussions, one can conclude that platoons of autonomous vehicles with constant spacing policy are incapable of handling common practices in managing transportation systems. It should be noted that since the behavior of nonlinear dynamic systems around the equilibrium states is similar to the corresponding linear system, the result can be extended to the nonlinear systems. Thus, no controller can guarantee the string stability in a large platoon based on the constant spacing policy.

## VI. CONCLUSION

Speed drops are common throughout the transportation systems to ensure safety and improve congestion. However, they can cause shockwave formation and breakdown. Accordingly, this paper investigated the possibility of reacting to such sudden speed drops within the platoons of autonomous vehicles controlled by constant spacing policy. It was analytically shown that it is indeed impossible to ensure strong stability, while following the speed profile with a sudden drop. One potential solution, that has been left for future research, is to refine the constant spacing policy to widen the stability region.

R<small>EFERENCES</small>